\input amstex
\documentstyle{conm-p}
\NoBlackBoxes

\issueinfo{00}
  {}
  {}
  {XXXX}

\topmatter
\title Inverse Scattering on the Line with Incomplete Scattering Data\endtitle
\author Tuncay Aktosun\endauthor
\leftheadtext{Tuncay Aktosun}

\address Department of Mathematics and Statistics, Mississippi State University,
Mississippi State, Mississippi 39762 \endaddress

\email aktosun\@math.msstate.edu\endemail

\thanks The research leading to this article
was supported in part by the National Science Foundation under
grant DMS-0204437 and the Department of Energy under grant
DE-FG02-01ER45951.\endthanks



\abstract The Schr\"odinger equation is considered on the line
when the potential is real valued, compactly supported, and square
integrable. The nonuniqueness is analyzed in the recovery of such
a potential from the data consisting of the ratio of a
corresponding reflection coefficient to the transmission
coefficient. It is shown that there are a discrete number of
potentials corresponding to the data
and that their $L^2$-norms are related to each other in a
simple manner. All those potentials are
identified, and it is shown how an additional estimate on
the $L^2$-norm in the data can uniquely identify the corresponding
potential. The recovery is illustrated with some explicit
examples.
\endabstract

\endtopmatter

\document

\head 1. Introduction \endhead

In this paper we analyze the recovery of the potential in the
Schr\"odinger equation on the line from a set of scattering data
containing no information on bound states. Our work is motivated
by the following question of Paul Sacks: Consider two potentials
in the Schr\"odinger equation where one potential is obtained from
the other by adding a bound state. Can we compare the $L^2$-norms
of these two potentials, and can we conclude that the potential
with fewer bound states has a smaller $L^2$-norm? By using (2.13)
and (2.17) these questions can be answered as follows: Take a
square-integrable potential and add a bound state with bound-state
energy $-\kappa^2$ and any bound-state norming constant. The new
potential will have a larger $L^2$-norm differing from the
previous $L^2$-norm by the exact value of $16\kappa^3/3.$ Note
that such a difference is independent of the value of the norming
constant used, and hence $L^2$-norms of square-integrable
potentials are affected only by bound-state energies and not by
norming constants.

Our work is also motivated by the work of Rundell and Sacks [1],
where it was shown that a bounded, real-valued,
compactly-supported potential with a sufficiently small $L^2$-norm
is uniquely determined by the corresponding ratio of a reflection
coefficient to the transmission coefficient. With the help of the
results in [2], our work here quantifies the smallness of the
$L^2$-norm in the result of [1]. In Section 3 we present the exact
least upper bound for that $L^2$-norm, below which we are assured
the unique determination of a real-valued, compactly-supported,
square-integrable potential in terms of the ratio of a reflection
coefficient to the transmission coefficient; we do not require the
potential to be bounded.

Let us now establish our notation. We consider the Schr\"odinger
equation
$$\psi''(k,x)+k^2\,\psi(k,x)=
V(x)\,\psi(k,x),\qquad x\in\bold R,\tag 1.1$$ where the potential
$V$ belongs to the Faddeev class, i.e. it is real valued,
measurable, and in $L^1_1(\bold R),$ the class of measurable
functions on the real axis $\bold R$ such that $\int_{-\infty}^\infty
dx\,(1+|x|)\,|V(x)|$ is finite. The prime is used for the
derivative with respect to the spatial coordinate $x.$ The Jost
solutions $f_{\text l}$ and $f_{\text r},$ from the left and
right, respectively, satisfy the respective boundary conditions
$$e^{-ikx}f_{\text l}(k,x)=1+o(1),\quad
e^{-ikx}f'_{\text l}(k,x)=ik+o(1),\qquad x\to+\infty,$$
$$e^{ikx}f_{\text r}(k,x)=1+o(1),\quad
e^{ikx}f'_{\text r}(k,x)=-ik+o(1),\qquad x\to-\infty,$$ and the
transmission coefficient $T,$ and the reflection coefficients $L$
and $R,$ from the left and right, respectively, are obtained from
the spatial asymptotics
$$f_{\text l}(k,x)=\displaystyle\frac{e^{ikx}}{T(k)}
+\displaystyle\frac{L(k)\,e^{-ikx}}{T(k)}+o(1), \qquad x\to-\infty,$$
$$f_{\text r}(k,x)=\displaystyle\frac{e^{-ikx}}{T(k)}
+\displaystyle\frac{R(k)\,e^{ikx}}{T(k)}+o(1), \qquad x\to+\infty.$$

A bound state of (1.1) is a square-integrable solution, and such
states occur only at the $k$-values on $\bold I^+$ in the upper
half complex $k$-plane ${\bold C^+}$ where $T(k)$ has (simple) poles.
Note that $\bold I^+:=i(0,+\infty)$ denotes the positive imaginary
axis. Later we will let ${\overline{\bold C^+}}:={\bold C^+}
\cup\bold R$ and $\bold
I^-:=i(-\infty,0).$ The behavior at $k=0$ tells us whether the
potential in (1.1) is generic or exceptional: The generic case
occurs if $T(0)=0$ and the exceptional case occurs if $T(0)\ne 0.$
For a review of scattering and bound states of (1.1), the reader
is referred to [3-9] and the references therein.

Our paper is organized as follows: In Section~2 we briefly review
the effect of adding a bound state to a potential and show that
certain integrals of the resulting potential remain unaffected by
the bound-state norming constant but affected only by the
bound-state energy and in a rather simple manner. In Section 3 we
analyze a consequence of the result of Section 2 in the recovery
of a real-valued, compactly-supported, square-integrable potential
in terms of the data $L(k)/T(k).$ We show that, corresponding to
that data, there are a discrete number of potentials, and an
additional estimate on the $L^2$-norm in the data allows the
unique identification of a potential among all. We also illustrate
the recovery with some explicit examples.

\head 2. Effect of Bound States on Norms of a Potential \endhead

Let $V^{[0]}$ denote a potential in the Faddeev class with no
bound states. We use $V^{[N]}$ for the potential obtained from
$V^{[0]}$ by adding $N$ bound states at $k=i\kappa_j$ with the
corresponding bound-state dependency constants $\gamma_j,$ where
we have the ordering $0<\kappa_1<\dots<\kappa_N.$ The superscript
$[j]$ refers to quantities associated with the potential
$V^{[j]};$ for example, $T^{[j]},$ $R^{[j]},$ and $L^{[j]}$ denote
the scattering coefficients, and $f^{[j]}_{\text l}$ and
$f^{[j]}_{\text r}$ denote the left and right Jost solutions.
Recall [4,9] that the dependency constants $\gamma_j$ are defined
as
$$\gamma_j:=\displaystyle\frac{f^{[N]}_{\text l}(i\kappa_j,x)}
{f^{[N]}_{\text r}(i\kappa_j,x)},\qquad 1\le j\le N,$$ and the
sign of $\gamma_j$ is such that $(-1)^{N-j}\gamma_j>0.$ It is
already known that
$$T^{[N]}(k)=T^{[0]}(k)
\displaystyle\prod_{j=1}^N \displaystyle\frac{k+i\kappa_j}
{k-i\kappa_j}, \tag 2.1$$
$$R^{[N]}(k)=(-1)^NR^{[0]}(k)
\displaystyle\prod_{j=1}^N \displaystyle\frac{k+i\kappa_j}
{k-i\kappa_j}, \quad
L^{[N]}(k)=(-1)^NL^{[0]}(k) \displaystyle\prod_{j=1}^N
\displaystyle\frac{k+i\kappa_j}{k-i\kappa_j}. \tag 2.2$$
For the known facts
listed in this section, we refer the reader to [4], where it is
shown that bound states can be added to a potential via the
Darboux transformation. We have
$$V^{[j]}(x)-V^{[j-1]}(x)=-2\,\mu'_j(x),
\qquad 1\le j\le N,\tag 2.3$$ where we have defined
$$\mu_j(x):=\displaystyle\frac{\chi'_j(x)}
{\chi_j(x)},\quad \chi_j(x):=f^{[j-1]}_{\text l}(i\kappa_j,x)
+|\gamma_j|\,f^{[j-1]}_{\text r}(i\kappa_j,x) .\tag 2.4$$ It is
known that $\chi_j(x)$ is continuous, strictly positive, and
differentiable. In fact, as seen from (2.4) we have
$$\mu'_j(x)=V^{[j-1]}(x)+\kappa_j^2-\mu_j(x)^2,
$$
and hence from (2.3) it follows that
$$V^{[j]}(x)+V^{[j-1]}(x)=2\left[\mu_j(x)^2-\kappa_j^2\right]
.\tag 2.5$$ Define
$$I_{j,n}(x):=\left[V^{[j]}(x)-V^{[j-1]}(x)\right]
\left[V^{[j]}(x)+V^{[j-1]}(x)\right]^n,\qquad n\ge 0, \ \ 1\le j
\le N.\tag 2.6$$

\proclaim {Theorem 2.1} Let $V^{[j]}$ be the potential
obtained from $V^{[0]}$ by adding bound states of energy
$-\kappa_1^2,\dots,-\kappa_j^2,$ and assume that $V^{[0]}$ belongs
to the Faddeev class without any bound states. We then have
$$\int_{-\infty}^\infty dx\,I_{j,n}(x)=(-1)^{n+1}2^{n+2}\kappa_j^{2n+1}
\displaystyle\frac{n!}{(2n+1)!!}, \qquad n\ge 0,\ \ 1\le j \le N,\tag 2.7$$
where $(2n+1)!!:=(1)(3)(5)\cdots (2n+1).$
\endproclaim

\demo{Proof}
Taking the $n$th power in (2.5) and expanding the
result, from (2.3) and (2.5) we get
$$I_{j,n}(x)=-2^{n+1}\displaystyle\frac{d}{dx}
\displaystyle\sum_{p=0}^n (-1)^{n-p}\kappa_j^{2(n-p)} \left(\matrix n\\
p\endmatrix\right) \displaystyle\frac{\mu_j(x)^{2p+1}}{2p+1} ,\tag 2.8$$
where $\left(\matrix n\\
p\endmatrix\right):=\displaystyle\frac{n!}{p!\,(n-p)!}$ is the binomial
coefficient. It is already known that
$$\mu_j(x)=\cases \kappa_j+o(1),\qquad x\to+\infty,\\
-\kappa_j+o(1),\qquad x\to-\infty.\endcases\tag 2.9$$ Integrating
(2.6) on $\bold R$ and using (2.9), we get
$$\int_{-\infty}^\infty dx\,I_{j,n}(x)=(-1)^{n+1}2^{n+2}\kappa_j^{2n+1}
\displaystyle\sum_{p=0}^n (-1)^{p} \left(\matrix n\\
p\endmatrix\right)\displaystyle\frac{1}{2p+1}.\tag 2.10$$ Note that the
summation in (2.10) can be evaluated explicitly with the help of
$$\displaystyle\sum_{p=0}^n (-1)^{p} \left(\matrix n\\
p\endmatrix\right)\displaystyle\frac{1}{2p+1}= \displaystyle\int_0^1
dx\,(1-x^2)^n=\displaystyle\frac{n!}{(2n+1)!!}, \qquad n\ge 0.\tag 2.11$$
Thus, using (2.11) in (2.10) we establish (2.7).
\quad\qed
\enddemo

The result in (2.7) is remarkable in the sense that even though
the integrand $I_{j,n}(x)$ depends on the bound-state data
$\{\kappa_p,\gamma_p\}_{p=1}^j,$ its integral given in (2.7) is
independent of the bound-state data, except for a rather simple
$\kappa_j$-dependence.

For $n=0$ and $n=1,$ respectively, from (2.7) we get
$$\int_{-\infty}^\infty dx\,\left[V^{[j]}(x)-V^{[j-1]}(x)\right]
=-4 \kappa_j,\qquad 1\le j \le N,\tag 2.12$$
$$\int_{-\infty}^\infty dx\,\left[V^{[j]}(x)^2-V^{[j-1]}(x)^2\right]
=\displaystyle\frac{16}{3}\, \kappa_j^3,\qquad 1\le j \le N.\tag 2.13$$ By
summing both sides in each of (2.12) and (2.13) over $j$, we get
the following:

\proclaim {Corollary 2.2} Let $V^{[0]}$ be a potential in
the Faddeev class with no bound states; add $N$ bound states with
energy $-\kappa_1^2,\dots,-\kappa_N^2,$ resulting in the potential
$V^{[N]}.$ We then have
$$\int_{-\infty}^\infty dx\,\left[V^{[N]}(x)-V^{[0]}(x)\right]
=-4\displaystyle\sum_{j=1}^N \kappa_j,\tag 2.16$$
$$\int_{-\infty}^\infty dx\,\left[V^{[N]}(x)^2-V^{[0]}(x)^2\right]
=\displaystyle\frac{16}{3}\displaystyle\sum_{j=1}^N \kappa_j^3.\tag 2.17$$
\endproclaim

Let us indicate some resemblance between the result in (2.7) and
the conserved quantities for an evolution equation that is exactly
solvable by the inverse scattering transform [12-14]. For example,
consider the time-evolution of the scattering data of (1.1) as
$T(k)\mapsto T(k),$ $L(k)\mapsto L(k)\,e^{-8ik^3t},$
$\kappa_j\mapsto\kappa_j,$ and $\gamma_j\mapsto\gamma_j
e^{-8\kappa_j^3 t}.$ The potential of (1.1) then evolves as
$V(x)\mapsto u(x,t),$ where $u(x,t)$ satisfies the initial-value
problem for the Korteweg-de Vries equation (KdV)
$$u_t-6uu_x+u_{xxx}=0,\qquad
x\in\bold R,\ \ t>0; \quad u(x,0)=V(x).$$ It is known [12-14] that
$\int_{-\infty}^\infty dx\,u(x,t),$ $\int_{-\infty}^\infty
dx\,u(x,t)^2,$ and an infinite number of other integrals are
independent of $t$ even though their integrands contain $t$
explicitly. Such quantities are known as the conserved quantities
for the KdV. Consider now, for example, (2.16) and (2.17), and let
us time evolve the potentials $V^{[0]}(x)$ and $V^{[N]}(x)$ to
obtain the corresponding solutions $u^{[0]}(x,t)$ and
$u^{[N]}(x,t)$ of the KdV. Due to the fact that the bound-state
energies $-\kappa_j^2$ remain unchanged during the time evolution
and that the right hand sides in (2.16) and (2.17) do not contain
the dependency constants $\gamma_j,$ we have
$$\int_{-\infty}^\infty dx\,\left[u^{[N]}(x,t)-u^{[0]}(x,t)\right]
=-4\displaystyle\sum_{j=1}^N \kappa_j,$$
$$\int_{-\infty}^\infty dx\,\left[u^{[N]}(x,t)^2-u^{[0]}(x,t)^2\right]
=\displaystyle\frac{16}{3}\displaystyle
\sum_{j=1}^N \kappa_j^3.$$ Other similar
conserved quantities for the KdV can be obtained with the help of
(2.7).

\head 3. Recovery of the Potential from $L(k)/T(k)$ \endhead

In [1] the recovery of a bounded, real-valued, compactly-supported
potential is considered in terms of the data $\Cal
D(k):=L(k)/T(k)$ known for $k\in\bold R.$ In the class of such
potentials corresponding to the same $\Cal D(k),$ it was shown
(cf. Theorem 2.3 of [1]) that there exists a positive constant $C$
such that if $V_1$ and $V_2$ are two potentials with $L^2$-norms
not exceeding $C$ then $V_1\equiv V_2.$ The uniqueness and the
reconstruction were obtained by transforming the problem into an
equivalent time-domain problem; however, the value of $C$ was left
unspecified. In this section, we show how the value of $C$ can be
specified.

Recently, we have analyzed [2] the recovery of the potential $V$
of (1.1) from $\Cal D(k)$ when $V$ belongs to the Faddeev class.
In this inverse problem, the construction of $V$ is equivalent to
the construction of the data
$\{L(k),N,\{\kappa_j\},\{\gamma_j\}\},$ where $L$ is the left
reflection coefficient, $N$ is the number of bound states, the set
$\{-\kappa_j^2\}_{j=1}^N$ corresponds to the bound-state energies,
and the set $\{\gamma_j\}_{j=1}^N$ corresponds to the bound-state
dependency constants. We have four cases to consider:

\roster
\item"(a)" No information is available on the support of
$V,$ and the only data available is $\Cal D(k).$
\item"(b)" In addition to
$\Cal D(k),$ it is known that the support of $V$ is confined to a
half line. In this case, there is no loss of generality in
assuming that $V\equiv 0$ for $x<0.$
\item"(c)" In addition to
$\Cal D(k),$ it is known that the support of $V$ is confined a
finite interval. In this case, there is no loss of generality in
assuming that $V\equiv 0$ for $x\notin[0,1].$
\item"(d)" In addition to
$\Cal D(k)$ and knowledge that $V\equiv 0$ for $x\notin[0,1],$ it
is known that $V$ is square integrable and some information
related to the $L^2$-norm is available. Such additional
information may be in the form of a positive constant $C$ which
acts as an upper bound on the $L^2$-norm.
\endroster

Let us consider the construction of $V$ or equivalently of
$\{L(k),N,\{\kappa_j\},\{\gamma_j\}\}$ in each of these four
cases. For the analysis in the first three cases we refer the
reader to [2] and give a brief summary below. Our results show
that in case (c), given $\Cal D(k)$ for $k\in\bold R,$ we are able to
determine all the corresponding potentials, there are a discrete
number of such potentials, the $L^2$-norm of each such potential
is readily evaluated with the help of (2.17), and appropriate
additional information on the $L^2$-norm enables us to further
restrict the set of potentials corresponding to $\Cal D(k).$ We
also explain how the constant $C$ in Theorem~2.3 of [1] arises:
That constant allows us to identify the potential with the
smallest $L^2$-norm among all those corresponding to the same
$\Cal D(k).$ By analyzing the inverse  problem stated in (d), we
show how to determine the precise values of $C$ that can be used
in [1].

\subhead {Case (a): Recovery of $V$ from $\Cal
D$ with no Support Information} \endsubhead

If no information other than $\Cal D(k)$ is available, we have the
following:

\roster
\itemitem"(a.i)" If $\Cal D(k)$ is bounded at $k=0,$ then there is
no restriction on $N$ and hence $N\in\{0,1,2,\dots\}.$ Note that
this case corresponds to the exceptional case for (1.1).
\itemitem"(a.ii)" If $\Cal D(k)$ is unbounded at $k=0,$
then $\lim_{k\to 0}[2ik\,\Cal D(k)]$ is either a positive constant
or a negative constant. Thus, either $\Cal D(k)\to -\infty$ or
$\Cal D(k)\to +\infty$ as $k\to 0$ on $\bold I^+.$ In the former
case $N$ must be even, i.e. $N\in\{0,2,4,\dots\};$ in the latter
case $N$ must be odd, i.e. $N\in\{1,3,5,\dots\}.$ Note that both
these correspond to the generic case for (1.1).
\itemitem"(a.iii)" For each $N$-value resulting from (i) or (ii),
given $\Cal D(k)$ there corresponds a $2N$-parameter family of
potentials where the parameter set is
$\{\kappa_j,\gamma_j\}_{j=1}^N.$ There are no restrictions on the
$\kappa_j$ other than $0<\kappa_1<\dots<\kappa_N.$ There are no
restrictions on the $\gamma_j$ other than $(-1)^{N-j}\gamma_j>0.$
\endroster

 From the data $\Cal D(k)$ known for
$k\in\bold R,$ one uniquely constructs
$$T^{[0]}(k)=\exp\left(\displaystyle
\frac{-1}{2\pi i}\int_{-\infty}^\infty
ds\,\displaystyle\frac{\log\left(1+|\Cal D(s)|^2\right)}
{s-k-i0^+}\right), \qquad k\in{\overline{\bold C^+}}.\tag 3.1$$
Then, with the help
of (2.1), it is seen that the set $\{\Cal D(k),N,\{\kappa_j\}\}$
leads to the left reflection coefficient given by
$$L(k)=\Cal D(k)\,T^{[0]}(k)\displaystyle\prod_{j=1}^N\displaystyle
\frac{k+i\kappa_j}{k-i\kappa_j},\qquad k\in\bold R.\tag 3.2$$
Note that $T^{[0]}(k)$ appearing in (3.1) and (3.2) corresponds to the
transmission coefficient for the potential $V^{[0]},$ which is
obtained by removing all the $N$ bound states from $V.$ The left
and right reflection coefficients, $L^{[0]}(k)$ and $R^{[0]}(k),$
respectively, corresponding to $V^{[0]}$ are uniquely determined
only in the generic case as [cf. (2.2)]
$$L^{[0]}(k)=(-1)^N\Cal D(k)\,T^{[0]}(k),\quad
R^{[0]}(k)=(-1)^{N-1}\Cal D(-k)\,T^{[0]}(k),
\qquad k\in\bold R,\tag 3.3$$
because only in the generic case $(-1)^N$ is uniquely
determined from $\Cal D(k).$ In the exceptional case, the value of
$(-1)^N$ cannot be determined from $\Cal D(k)$ and hence there are
two choices for $V^{[0]},$ which we denote by $V_1^{[0]}$ and
$V_2^{[0]},$ respectively, with the corresponding scattering
coefficients determined in terms of $\Cal D(k)$ and $T^{[0]}(k)$
in (3.1) as follows:
$$T_1^{[0]}(k)=T^{[0]}(k),
\quad L_1^{[0]}(k)=\Cal D(k)\,T^{[0]}(k),\quad R_1^{[0]}(k)=-\Cal
D(-k)\,T^{[0]}(k),$$
$$T_2^{[0]}(k)=T^{[0]}(k),
\quad L_2^{[0]}(k)=-\Cal D(k)\,T^{[0]}(k),\quad R_2^{[0]}(k)=\Cal
D(-k)\,T^{[0]}(k).$$ For the comparison of two potentials with the
same transmission coefficient but with reflection coefficients
differing in sign, the reader is referred to [9-11]. As the next
proposition shows, even though $V_1^{[0]}\not\equiv V_2^{[0]},$
some of their characteristic features are related.

\proclaim {Proposition 3.1} Let $V_1^{[0]}$ and
$V_2^{[0]}$ be two exceptional potentials in the Faddeev class
with no bound states, and assume that $T_1^{[0]}\equiv T_2^{[0]},$
$L_1^{[0]}\equiv - L_2^{[0]},$ and $R_1^{[0]}\equiv - R_2^{[0]},$
i.e. their reflection coefficients differ in sign and their
transmission coefficients are the same for all $k\in\bold R.$ Then we
have the following:

\roster
\item"(i)" $\displaystyle\int_{-\infty}^\infty dx\,\left[V_2^{[0]}(x)-V_1^{[0]}(x)\right]
\left[V_2^{[0]}(x)+V_1^{[0]}(x)\right]^n=0,\qquad n\ge 0.$
\item"(ii)" $V_1^{[0]}$ vanishes on a half line
if and only if $V_2^{[0]}$ vanishes on the same half line.
Consequently, $V_1^{[0]}$ vanishes outside some interval if and
only if $V_2^{[0]}$ vanishes outside that interval.
\item"(iii)" If $V_1^{[0]}$ vanishes on
$\bold R^-$ and is continuous on the interval $(0,\delta)$ for
some $\delta>0,$ then $V_1^{[0]}(0^+)=-V_2^{[0]}(0^+).$ Similarly,
if $V_1^{[0]}$ vanishes on $\bold R^+$ and is continuous on the
interval $(-\delta,0)$ for some $\delta>0,$ then
$V_1^{[0]}(0^-)=-V_2^{[0]}(0^-).$
\endroster

\endproclaim

\demo {Proof} From (2.24) and (2.25) of [11] we have
$$V_2^{[0]}(x)=V_1^{[0]}(x)-2\,\rho'_1(x)=
2\,\rho_1(x)^2-V_1^{[0]}(x),\tag 3.4$$ where
$$\rho_1(x):=\displaystyle\frac{f^{[0]\prime}_{1{\text l}}(0,x)}{
f^{[0]}_{1{\text l}}(0,x)}=\displaystyle\frac{f^{[0]\prime}_{1{\text
r}}(0,x)}{ f^{[0]}_{1{\text r}}(0,x)},$$ with $f^{[0]}_{1{\text
l}}(k,x)$ and $f^{[0]}_{1{\text r}}(k,x)$ being the left and right
Jost solutions for the potential $V_1^{[0]}.$ It is known that
$f^{[0]}_{1{\text l}}(0,x)$ is continuous and strictly positive
and that $\rho_1(x)=o(1/x)$ as $x\to\pm\infty.$ Hence, with the
help of (3.4) we get [cf. (2.8)]
$$\left[V_2^{[0]}(x)-V_1^{[0]}(x)\right]
\left[V_2^{[0]}(x)+V_1^{[0]}(x)\right]^n
=-\displaystyle\frac{2^{n+1}}{2n+1}\left[ \rho_1(x)^{2n+1}\right]',\qquad
n\ge 0,$$ and integrating both sides over $\bold R$ we obtain (i). To
prove (ii), notice that there is no loss of generality in choosing
the half line as $\bold R^-.$ If $V_1^{[0]}\equiv 0$ for $x<0,$ then
$f^{[0]}_{1{\text r}}(0,x)=1$ for $x\le 0,$ and hence
$\rho_1(x)=0$ for $x\le 0.$ Thus, from (3.4) it follows that
$V_2^{[0]}\equiv 0$ for $x<0$ as well. Conversely, it follows that
$V_1^{[0]}\equiv 0$ on $\bold R^-$ whenever $V_2^{[0]}\equiv 0$ there;
thus, we have proved (ii). From (3.4) we see that the first
statement in (iii) holds whenever $\rho_1(0^+)=0,$ which is the
case due to the continuity of $\rho_1(x)$ at $x=0$ and
$\rho_1(x)=0$ for $x\le 0,$ which is satisfied when
$V_1^{[0]}\equiv 0$ for $x<0.$ The second statement in (iii) is
obtained in a similar manner.
\quad\qed
\enddemo

Note that Proposition~3.1(i) holds even when $n$ is a noninteger.
By letting $n=0$ and $n=1$ there and using the fact that a
potential in the Faddeev class is integrable, we obtain
$$\displaystyle\int_{-\infty}^\infty dx\,
V_1^{[0]}(x)=\displaystyle\int_{-\infty}^\infty dx\, V_2^{[0]}(x),\qquad
\displaystyle\int_{-\infty}^\infty dx\, \left[V_1^{[0]}(x)^2-
V_2^{[0]}(x)^2\right]=0.\tag 3.5$$ For smooth potentials, we refer
the reader to (2.101) of [10] for results similar to (3.5) and
their generalizations.

\subhead {Case (b): Recovery of $V$ from $\Cal
D$ with Half-line Support} \endsubhead

If $\Cal D(k)$ is given for $k\in\bold R$ and if it is also known that
$V\equiv 0$ for $x<0,$ then, in addition to all the results given
in case (a), in particular, in addition to (a.i) and (a.ii), we
have the following improvements:

\roster
\itemitem"(b.iii)" $\Cal D(k)$ has a unique analytic extension
to $k\in{\bold C^+}$ and such an extension is uniquely determined by our
data $\Cal D(k)$ known for $k\in\bold R.$ The value of $N$ must
satisfy $N\le Z+1,$ where $Z$ denotes the number of zeros of $\Cal
D(k)$ on $\bold I^+.$ In fact, from the proof of Proposition~3.1
in [2] it follows that if $\Cal D(k)$ has multiple zeros on $\bold
I^+,$ then $Z$ is actually the number of distinct zeros of odd
multiplicity, without counting the multiplicities.
\itemitem"(b.iv)" For each $N$-value resulting from restrictions
(a.i), (a.ii), and (b.iii), given $\Cal D(k)$ for $k\in\bold R,$ there
corresponds an $N$-parameter family of potentials where the
parameter set is $\{\kappa_j\}_{j=1}^N.$ The $\kappa_j$ satisfy
the restrictions $0<\kappa_1<\dots<\kappa_N$ and $(-1)^{N-j}\Cal
D(i\kappa_j)>0$ for $j=1,\dots,N.$ The latter restriction confines
the $\kappa_j$ to subintervals whose endpoints are uniquely
determined by the zeros of $\Cal D(k)$ on $\bold I^+.$ The
dependency constants $\gamma_j$ are uniquely determined as
$\gamma_j=\Cal D(i\kappa_j)$ and hence they are not free
parameters. The left reflection coefficient $L(k)$ given in (3.2)
becomes meromorphic in ${\bold C^+}$ with simple poles at $k=i\kappa_j$
for $j=1,\dots,N.$ Thus, (3.2) now holds for $k\in{\overline{\bold C^+}}$ and the
left reflection coefficient $L^{[0]}(k)$ given in (3.3) becomes
analytic in ${\bold C^+}.$
\endroster

\subhead {\bf Case (c): Recovery of $V$ from $\Cal
D$ with Compact Support} \endsubhead

If $\Cal D(k)$ for $k\in\bold R$ is given and if it is also known that
$V\equiv 0$ for $x\notin[0,1],$ then, in addition to all the
results in cases (a) and (b), in particular, in addition to (a.i)
and (a.ii), we have the following improvements:

\roster
\itemitem"(c.iii)" The quantity
$k\,\Cal D(k)$ has a unique analytic extension to the entire
complex plane, and such an extension is uniquely determined by our
data $\Cal D(k)$ known for $k\in\bold R.$ Moreover, as in (b.iii) the
value of $N$ must satisfy $N\le Z+1,$ where $Z$ is the number of
zeros of $\Cal D(k)$ on $\bold I^+$ having odd multiplicities,
without counting the multiplicities.
\itemitem"(c.iv)" For each $N$-value resulting from restrictions
(a.i), (a.ii), and (c.iii), given $\Cal D(k)$ for $k\in\bold R,$ there
correspond a discrete number of potentials where the discrete
parameter set is $\{\kappa_j\}_{j=1}^N.$ The set
$\{\kappa_j\}_{j=1}^N$ must be a subset of $\{\beta_m\}$ and
satisfy the additional restrictions $0<\kappa_1<\dots<\kappa_N$
and $(-1)^{N-j}\Cal D(i\kappa_j)>0$ for $j=1,\dots,N.$ Here, each
$k=-i\beta_m$ corresponds to a zero of $1/T^{[0]}(k)$ on $\bold
I^-,$ where $T^{[0]}(k)$ is the quantity in (3.1), and
$k/T^{[0]}(k)$ is now entire on $\bold C$ and uniquely constructed
via (3.1) from our data $\Cal D(k)$ known for $k\in\bold R.$ The
values $k=-i\beta_m$ correspond to the (real) resonances of
$V^{[0]}.$ For an answer to the question  whether the set
$\{\beta_m\}$ is a finite set or an infinite set, we refer the
reader to [15]. Informally speaking, if $V^{[0]} \in
C^\infty_0[0,1]$ and the order of the zero of $V^{[0]}$ at $x=0$
or at $x=1$ is infinite, then the set $\{\beta_m\}$ may be
infinite; otherwise, it is a finite set.
\endroster

\subhead {\bf Case (d): Recovery of $V$ from $\Cal
D$ with Compact Support and $L^2$-norm} \endsubhead

Let us assume that $\Cal D(k)$ is given for $k\in\bold R$ and it is
known that $V\equiv 0$ for $x\notin[0,1],$ $V\in L^2[0,1],$ and
$||V||\le C,$ where we denote the $L^2$-norm of $V$ as
$||V||:=\sqrt{\int_{-\infty}^\infty dx\, V(x)^2}.$ We will
determine the precise values of $C$ that assure a unique or
nonunique determination of $V$ from $\Cal D.$

As outlined below (3.3) in case (a), given $\Cal D(k)$ for
$k\in\bold R,$ we are able to uniquely determine $V^{[0]}$ when $\Cal
D$ is singular at $k=0,$ and we determine two distinct potentials
$V_1^{[0]}$ and $V_2^{[0]}$ if $\Cal D$ is finite at $k=0.$ In the
latter case, we know from (3.5) that
$||V_1^{[0]}||=||V_2^{[0]}||.$ Thus, $\Cal D(k)$ uniquely
determines the $L^2$-norm of $V^{[0]},$ even though there are two
distinct choices for $V^{[0]}$ in the exceptional case. Let us
denote that unique value by $||V^{[0]}||.$

As seen from (c.iv), for each allowed allowed integer $N,$ $\Cal
D(k)$ uniquely [2] determines a discrete number of ordered sets
$\{\kappa_j\}_{j=1}^N$ with the ordering
$0<\kappa_1<\dots<\kappa_N$ related to the bound states of
$V^{[N]}.$ Let us define
$$C_0:=||V^{[0]}||;
\quad C_N:=\left[||V^{[0]}||^2+\displaystyle\frac{16}{3}\displaystyle\sum _{j=1}^N
\kappa_j^3\right]^{1/2}.$$ Thus, for each $N,$ $C_N$ consists of a
sequence of values. Clearly, $C_0$ consists of a single number. By
listing all the elements in $C_N$ for all allowed $N$-values, we
obtain a discrete set of ordered positive numbers consisting of
various $\kappa_j$ values, and we denote that discrete set by
$\{C_N\}.$ This set is a subset of $\{\beta_m\},$ as indicated in
(c.iv). The smallest number in the ordered set  $\{C_N\}$ is
strictly less than the next larger number due to the fact that
each set $\{\kappa_j\}_{j=1}^N$ with the largest allowable $N$
consists of distinct positive elements. This allows us to
determine the value of $C$ in the inequality $||V||\le C$ in order
to determine a unique potential $V$ corresponding to our data
$\Cal D.$ By choosing $C$ as greater than or equal to the smallest
number in the set $\{C_N\}$ but strictly less than the next larger
element, we will uniquely determine the potential $V.$ Next we
illustrate this determination with some explicit examples.

As our scattering data let us use $\Cal D(k)=\displaystyle\frac{-\epsilon
\,e^{ik}\sin\sqrt{k^2+\epsilon}} {2ik\,\sqrt{k^2+\epsilon}},$
where $\epsilon$ is a positive parameter. In fact, one
corresponding potential is the square well of depth $\epsilon$
supported on the interval $[0,1].$ For each value of $\epsilon,$
let us obtain all the potentials corresponding to $\Cal D(k)$ with
support confined to $[0,1]$ and specify their $L^2$-norms. We have
$\lim_{k\to 0}[2ik\,\Cal
D(k)]=-\sqrt{\epsilon}\,\sin\sqrt{\epsilon},$ and hence the
exceptional case occurs when $\sqrt{\epsilon}/\pi$ is a positive
integer and the generic case occurs otherwise. The zeros of $\Cal
D(k)$ on $\bold I^+$ occur when $\sin\sqrt{k^2+\epsilon}=0,$ and
hence these are all simple zeros occurring at
$k=i\sqrt{\epsilon-(j-1)^2\pi^2}$ for $j=1,\dots,Z,$ with $Z$
being equal to $\lfloor \sqrt{\epsilon}/\pi\rfloor,$ i.e. the
greatest integer less than or equal to $\sqrt{\epsilon}/\pi.$ As
$k\to\infty$ on $\bold I^+,$ we have $\Cal D(k)\to 0^+.$ As $k\to
0$ on $\bold I^+,$ we get $(-1)^Z\Cal D(k)\to 0^+$ in the
exceptional case, and $(-1)^Z\Cal D(k)\to +\infty$ in the generic
case. Define
$$\displaystyle\frac{1}{\tau(k)}:=e^{ik}\left[\cos \sqrt{k^2+\epsilon}
+\displaystyle\frac{2k^2+\epsilon}{2ik\sqrt{k^2+\epsilon}}
\,\sin\sqrt{k^2+\epsilon}\right].$$
Note that $\tau(k)$
corresponds to the transmission coefficient of the square-well
potential of depth $\epsilon$ supported on $[0,1].$ It is known
that $1/\tau(k)$ has exactly $Z+1$ (simple) zeros on $\bold I^+,$
which we denote by $\xi_j$ with the ordering
$0<\xi_1<\dots<\xi_{Z+1}.$ The quantity in (3.1) is obtained as
$$\displaystyle\frac{1}{T^{[0]}(k)}
=\displaystyle\frac{1}{\tau(k)}
\displaystyle\prod_{j=1}^{Z+1}\displaystyle\frac
{k+i\xi_j}{k-i\xi_j}.\tag 3.6$$

\example {Example 3.1} When $\epsilon=5,$ we are in the
generic case and $Z=0.$ Hence, $N\le 1,$ but $\Cal D(k)\to
+\infty$ on $\bold I^+$ indicates that $N$ must be odd. Thus,
$N=1$ is the only allowed value. In this case, $1/T^{[0]}(k)$
given in (3.6) has two zeros on $\bold I^-$ at $k=-i\beta_j$ with
$\beta_1=1.5433\overline{4}$ and $\beta_2=1.585\overline{7}.$ We
use an overline to indicate roundoff. In (3.6) we have
$\xi_1=\beta_2.$ Corresponding to $\Cal D(k)$ we have two
potentials $V_1^{[1]}$ and $V_2^{[1]},$ having bound states at
$k=i\beta_1$ and $k=i\beta_2,$ respectively. Note that $V_2^{[1]}$
is the square well of depth $\epsilon.$ We have
$||V_1^{[1]}||=4.8312\overline{6}$ and $||V_2^{[1]}||=5.$ Thus,
knowledge of any $C$ satisfying $||V_1^{[1]}||\le C<||V_2^{[1]}||$
helps us to identify $V_1^{[1]}$ or $V_2^{[1]}$ as the unique
potential corresponding to $\Cal D(k).$ The left reflection
coefficients $L_1^{[1]}$ and $L_2^{[1]}$ corresponding to
$V_1^{[1]}$ and $V_2^{[1]},$ respectively,
are obtained from (3.2) as
$$L_j^{[1]}(k)=\Cal D(k)\,T^{[0]}(k)\,\displaystyle
\frac{k+i\beta_j}{k-i\beta_j},\qquad j=1,2.$$
Note that $V_1^{[1]}$ and $V_2^{[1]}$ can uniquely
be constructed [11] from $L_1^{[1]}$ and $L_2^{[1]},$
respectively, because they vanish for $x<0.$
\endexample

\example {Example 3.2} When $\epsilon=\pi^2,$ we are in the
exceptional case and $Z=0.$ Hence, both $N=0$ and $N=1$ are
allowed. In this case $1/T^{[0]}(k)$ given in (3.6) has only one
zero on $\bold I^-$ at $k=-i\beta_1$ with
$\beta_1=2.52258\overline{8}.$ Thus, in (3.6) we have
$\xi_1=\beta_1.$ Corresponding to $\Cal D(k)$ we have two
potentials $V^{[0]}$ and $V^{[1]},$ the former with no bound
states and the latter with one bound state at $k=i\beta_1.$ Note
that $V^{[1]}$ is the square well of depth $\epsilon.$ We have
$||V^{[0]}||=3.3853\overline{7}$ and $||V^{[1]}||=\pi^2.$ Thus,
knowledge of any $C$ satisfying $||V^{[0]}||\le C<||V^{[1]}||$
helps us to identify either $V^{[0]}$ or $V^{[1]}$ as the unique
potential corresponding to $\Cal D(k).$ The left reflection
coefficients $L^{[0]}$ and $L^{[1]}$ corresponding to $V^{[0]}$
and $V^{[1]},$ respectively, are obtained from (3.2) as
$$L^{[0]}(k)=\Cal D(k)\,T^{[0]}(k),\quad
L^{[1]}(k)=\Cal D(k)\,T^{[0]}(k)\,
\displaystyle\frac{k+i\beta_1}{k-i\beta_1}.$$
Having
$L^{[0]}$ and $L^{[1]}$ at hand, the potentials $V^{[0]}$ and
$V^{[1]}$ can be uniquely constructed.
\endexample

\example {Example 3.3} When $\epsilon=20,$ we are in the
generic case and $Z=1.$ Hence, $N\le 2,$ but $\Cal D(k)\to
-\infty$ on $\bold I^+$ indicates that $N$ must be even. Thus,
$N=0$ and $N=2$ are the only possibilities. In this case
$1/T^{[0]}(k)$ given in (3.6) has two zeros  on $\bold I^-$ at
$k=-i\beta_1$ with $\beta_1=\xi_1=1.9302\overline{1}$ and
$k=-i\beta_2$ with $\beta_2=\xi_2=3.9255\overline{6}.$ When $N=2,$
the only potential $V^{[2]}$ corresponding to $\Cal D(k)$ is the
square well of depth $\epsilon$ with support $[0,1].$ When $N=0,$
the corresponding potential $V^{[0]}$ is uniquely determined from
$\Cal D(k)$ and its left reflection coefficient $L^{[0]}(k)$ is
obtained from (3.2) as
$$L^{[0]}(k)=\Cal D(k)\,T^{[0]}(k)\,\displaystyle\frac
{(k-i\beta_1)(k-i\beta_2)}{(k+i\beta_1)(k+i\beta_2)}.$$ In this
case we have $||V^{[0]}||=6.2463\overline{5}$ and
$||V^{[2]}||=20.$ Thus, an appropriate specification of the upper
limit on the $L^2$-norm of the potential allows the unique
identification of $V^{[0]}$ or $V^{[2]}$ from $\Cal D(k).$
\endexample

\example {Example 3.4} When $\epsilon=130,$ the allowed
values for $N$ are $0,$ $2,$ and $4.$ In this case $1/T^{[0]}(k)$
given in (3.6) has six zeros on $\bold I^-$ at $k=-i\beta_j$ with
$\beta_1=4.8729\overline{5},$ $\beta_2=8.2260\overline{7},$
$\beta_3=8.3286\overline{5},$ $\beta_4=10.087\overline{9},$
$\beta_5=10.740\overline{7},$ $\beta_6=11.08\overline{5}.$ For
$N=0,$ the only potential corresponding to $\Cal D(k)$ has norm
$||V^{[0]}||=23.96\overline{8}.$ For $N=2,$ there are five
potentials  corresponding to $\Cal D(k)$ with norms
$||V_1^{[2]}||=64.50\overline{9},$
$||V_2^{[2]}||=65.366\overline{8},$
$||V_3^{[2]}||=91.956\overline{6},$
$||V_4^{[2]}||=115.38\overline{7},$
$||V_5^{[2]}||=120.19\overline{7},$ where $V_1^{[2]}$ has bound
states $\{-\beta_1^2,-\beta_2^2\},$ $V_2^{[2]}$ has
$\{-\beta_1^2,-\beta_3^2\},$ $V_3^{[2]}$ has
$\{-\beta_1^2,-\beta_6^2\},$ $V_4^{[2]}$ has
$\{-\beta_4^2,-\beta_6^2\},$ and $V_5^{[2]}$ has
$\{-\beta_5^2,-\beta_6^2\}.$ For $N=4,$ there are four potentials
corresponding to $\Cal D(k)$ with norms $||V_1^{[4]}||=130,$
$||V_2^{[4]}||=130.43\overline{2},$
$||V_3^{[4]}||=134.28\overline{7},$
$||V_4^{[4]}||=134.70\overline{5},$ where $V_1^{[4]}$ has bound
states $\{-\beta_1^2,-\beta_2^2,-\beta_4^2,-\beta_6^2\},$
$V_2^{[4]}$ has $\{-\beta_1^2,-\beta_3^2,-\beta_4^2,-\beta_6^2\},$
$V_3^{[4]}$ has $\{-\beta_1^2,-\beta_2^2,-\beta_5^2,-\beta_6^2\},$
and finally $V_4^{[4]}$ has
$\{-\beta_1^2,-\beta_3^2,-\beta_5^2,-\beta_6^2\}.$
Thus, some appropriate knowledge on the $L^2$-norm of the
potential allows us to pick a unique potential among all these
$16$ potentials corresponding to the same $\Cal D(k).$ Note that
$V_1^{[4]}$ is the square well of depth $\epsilon.$
\endexample

\refstyle{N}

\enddocument